\documentclass[11pt,a4paper]{article}
\pdfoutput=1
\usepackage{jheppub}

\usepackage{multirow, graphicx,amssymb,url,mathrsfs,amsmath}
\usepackage{wrapfig,boxedminipage,epsfig}
\usepackage{amsxtra,amstext,latexsym,dsfont,amsfonts}
\usepackage{color}
\usepackage[dvipsnames]{xcolor}
\usepackage{subfig}
\usepackage{slashed,comment}
\usepackage{contour}
\usepackage{tikz}
\usetikzlibrary{calc,patterns,angles,quotes}
\usetikzlibrary{decorations.pathreplacing,decorations.markings,snakes}
\usepackage{tabularx, array}
\usepackage{mathtools}
\usepackage{eufrak,booktabs}
\usetikzlibrary{shapes.geometric}
\usetikzlibrary{shapes,backgrounds}

\def\D  {\Delta}       
       
\renewcommand{\a}{\alpha}

      \renewcommand{\t}{\tau}

\newcommand{\bra}[1]{\mbox{$\langle #1 |$}}
\newcommand{\ket}[1]{\mbox{$| #1 \rangle$}}

\newcommand{\braket}[2]{\mbox{$\langle #1  | #2 \rangle$}}

\newcommand{\brac}[1]{\langle #1 \rangle}

\newcommand{\Tr}{{\rm Tr}\,}

\renewcommand{\Re}{{ \rm Re}}
\renewcommand{\Im}{{ \rm Im}}

\newcolumntype{L}[1]{>{\raggedright\arraybackslash}p{#1}}
\newcolumntype{C}[1]{>{\centering\arraybackslash}p{#1}}
\newcolumntype{R}[1]{>{\raggedleft\arraybackslash}p{#1}}

\usepackage{xparse}

\title{Toward Krylov-based holography in double-scaled SYK}

\author[a]{Yichao Fu,}
\author[b,c,d]{Hyun-Sik Jeong,}
\author[a,e]{Keun-Young Kim,}
\author[b]{Juan F. Pedraza}
 
\emailAdd{yichao.fu@gm.gist.ac.kr}
\emailAdd{hyunsik.jeong@apctp.org}
\emailAdd{fortoe@gist.ac.kr}
\emailAdd{j.pedraza@csic.es}

\preprint{\texttt{IFT-UAM/CSIC-25-105, APCTP Pre2025 - 020}}

\affiliation[a]{Department of Physics and Photon Science, Gwangju Institute of Science and Technology, \\
123 Cheomdan-gwagiro, Gwangju 61005, Korea}
\affiliation[b]{Instituto de F\'isica Te\'orica UAM/CSIC, Calle Nicol\'as Cabrera 13-15, 28049 Madrid, Spain}
\affiliation[c]{Asia Pacific Center for Theoretical Physics, Pohang 37673, Korea}
\affiliation[d]{Department of Physics, Pohang University of Science and Technology, Pohang 37673, Korea}
\affiliation[e]{Research Center for Photon Science Technology, Gwangju Institute of Science and Technology, \\
123 Cheomdan-gwagiro, Gwangju 61005, Korea}

\abstract{Building on the duality between Krylov complexity and geodesic length in Jackiw-Teitelboim and sine-dilaton gravity, we develop a precise holographic dictionary for quantities in the Krylov subspace of the double-scaled Sachdev-Ye-Kitaev model (DSSYK). First, we demonstrate that the growth rate of Krylov state complexity corresponds to the wormhole velocity, and show that its expectation value in coherent states serves as a boundary diagnostic of firewall-like structures via bulk reconstruction. We also delineate an alternative bulk description in terms of the proper momentum of an infalling particle at early times, establishing a threefold duality between the Krylov complexity growth rate, wormhole velocity, and proper momentum, with clear regimes of validity. Beyond the first moments, we argue that higher-order Krylov complexities capture connected bulk contributions encoded by replica wormholes, while the logarithmic variant probes the replica saddle structure. Finally, within a third-quantized setting incorporating baby universes, we show that the Krylov entropy equals the von Neumann entropy of the parent-geometry density matrix obtained after tracing out baby universes, thereby quantifying information flow into the baby universe sector. Together, these results elevate Krylov-space observables to sharp probes of bulk dynamics and topology in ensemble-averaged 2D gravity.}

\begin{document}
\maketitle

%
\section{Introduction}\label{}
In recent years, Krylov complexity has emerged as a powerful tool for quantitatively characterizing operator growth and state evolution in quantum systems \cite{Parker:2018yvk,Balasubramanian:2022tpr,Caputa:2024vrn}. Remarkably, this quantity exhibits universal features in chaotic systems, although exceptions have been observed, particularly in systems dominated by unstable saddles \cite{Bhattacharjee:2022vlt, Huh:2023jxt, Aguilar-Gutierrez:2025hbf}. Nevertheless, Krylov complexity is widely considered one of the primary diagnostics of quantum chaos, in line with traditional indicators such as spectral statistics \cite{Baggioli:2024wbz, Alishahiha:2024vbf,Huh:2024ytz,Baggioli:2025knt}. For recent, comprehensive reviews of Krylov complexity, including further references, see~\cite{Nandy:2024evd,Rabinovici:2025otw}.

A variant of Krylov complexity, referred to as Krylov \emph{state} complexity \cite{Balasubramanian:2022tpr} (or spread complexity), tracks how a quantum state spreads in the Krylov subspace generated by a given Hamiltonian, via the Lanczos algorithm. This notion provides valuable insights into the system's time evolution, making it conceptually akin to computational complexity in holography, where the AdS/CFT correspondence~\cite{Maldacena:1997re, Witten:1998qj, Aharony:1999ti} relates the quantum dynamics of the dual theory to geometric quantities in the bulk. In particular, holographic complexity proposals such as ``complexity=volume'' (CV)~\cite{Stanford:2014jda}, ``complexity=volume 2.0'' (CV2.0)~\cite{Couch:2016exn}, ``complexity=action'' (CA)~\cite{Brown:2015bva}, and the broader ``complexity=anything'' (CAny) variants~\cite{Belin:2021bga,Belin:2022xmt} attempt to quantify the same intuitive idea: how `difficult' it is to construct a given boundary state using bulk gravitational observables.

This conceptual parallel naturally raises the question: \textit{can Krylov complexity be identified with—or serve as a boundary dual to—any of these gravitational complexity proposals?} Addressing this question is challenging in full generality, so it is instructive to begin with tractable toy models. Recent studies have examined holographic notions of complexity in Jackiw-Teitelboim (JT) gravity \cite{Brown:2018bms,Alishahiha:2018swh,Bhattacharya:2023drv,Fu:2024vin,Caceres:2025myu, Miyaji:2025yvm, Miyaji:2025jxy} and in variants of the Sachdev-Ye-Kitaev (SYK) model \cite{Chapman:2024pdw,Bhattacharjee:2022ave,Jha:2024nbl,Xu:2024gfm}, and have uncovered intriguing connections between the two, namely: (i) a duality between wormhole length and Krylov state complexity \cite{Rabinovici:2023yex,Aguilar-Gutierrez:2025pqp, Aguilar-Gutierrez:2025sqh}; and (ii) holographic realizations of Krylov operator complexity \cite{Ambrosini:2025hvo, Ambrosini:2024sre,Bhattacharyya:2025gvd,Kar:2021nbm,Jian:2020qpp,Aguilar-Gutierrez:2025mxf,Jeong:2026iac}. These setups, however, are limited insofar as JT gravity is dual only to certain triple- or double-scaled limits of SYK. To move beyond this, a broader duality between sine-dilaton gravity and DSSYK at finite $q$ has been proposed \cite{Blommaert:2024ymv,Blommaert:2024whf,Bossi:2024ffa}, within which a correspondence between CV (geodesic lengths) and Krylov state complexity has been confirmed to hold~\cite{Heller:2024ldz}. Here, our goal is to explore further dualities within the same framework, involving additional Krylov-subspace quantities.\footnote{See also~\cite{Jeong:2024jjn,Jeong:2025jyx} for an alternative approach in which Krylov complexity is evaluated directly in the bulk—specifically for near-horizon perturbations using the brickwall model~\cite{tHooft:1984kcu,Das:2022evy,Das:2023ulz}—rather than on the boundary. This offers a complementary route to diagnosing quantum chaotic dynamics of black holes.
}

In particular, we aim to explicitly identify the holographic dual of the \textit{growth rate of Krylov complexity} and to offer physical interpretations. On the quantum theory side, although the growth rate of spread complexity does not, by itself, carry decisive physical significance, it is known that the operator complexity problem with a matter insertion in the thermofield double (TFD) state of DSSYK can be mapped to a state complexity problem in the doubled Hilbert space evolved with the Hamiltonian $H_R-H_L$ \cite{Lin:2022rbf, Ambrosini:2024sre}. The growth rate of state complexity is therefore interesting in this setting, given its connection to the operator complexity growth rate, which is closely linked to Lyapunov exponents. On the gravity side, the growth rate of holographic complexity is crucial because the complexity itself is scheme-dependent (i.e., it requires renormalization), whereas the growth rate remains well defined. This rate is directly related to black hole thermodynamics and to the bulk symplectic form. In this work, we propose that this growth rate is dual to the wormhole velocity in the bulk geometry. Notably, previous studies have suggested an alternative interpretation, namely, that the same boundary quantity (its growth rate) corresponds to the proper momentum of an infalling particle in the bulk~\cite{Caputa:2024sux,Fan:2024iop,He:2024pox}. This prompts a natural question: \textit{Is the momentum interpretation the only valid dual, and under what circumstances might it fail?} One of our goals is to address this question within an analytically tractable framework consistent with the CV conjecture.

We further extend our analysis to the holographic duals of higher-order Krylov complexities \cite{Fan:2023ohh,Fu:2024fdm,Camargo:2024rrj}, the recently introduced logarithmic Krylov complexity \cite{Camargo:2026szl}, and Krylov entropy. Higher-order Krylov complexities are related to higher moments of the average position along the Krylov chain. In chaotic systems, they have been shown to exhibit clearer signatures of chaos—namely, a characteristic pre-saturation peak. The logarithmic Krylov complexity is proposed as a resolution of saddle-dominated scrambling and is defined analytically via the replica trick \cite{Camargo:2026szl}. In this sense, the higher-order complexities are R\'enyi-like quantities and, in the bulk, may receive contributions from replica-wormhole saddles. In particular, we uncover a gravitational dual of Krylov entropy and discuss its physical interpretation, especially in the context of a third-quantized bulk with baby universes \cite{Giddings:1988wv,Marolf:2020xie,Penington:2023dql}. Our discussion applies equally to both sine-dilaton and JT gravity, as both represent ensemble-averaged holographic theories.

The remainder of this paper is organized as follows.
In Section~\ref{sec-intro}, we review the relevant background on the DSSYK model, sine-dilaton gravity, and Krylov complexity.
In Section~\ref{sec-main}, we revisit the duality between Krylov complexity and wormhole length, and then propose a correspondence between the growth rate of Krylov complexity and wormhole velocity. Using expectation values in coherent states, we further interpret this quantity as a potential boundary signature of a firewall, based on bulk reconstruction arguments. We next analyze the alternative dual description in terms of the bulk proper momentum and clarify its limitations. Subsequently, we explore the gravitational duals of higher-order and logarithmic Krylov complexities, emphasizing their relation to replica wormholes. Finally, we present a proposal for a bulk dual of Krylov entropy and discuss its interpretation within the third-quantization framework.
In Section~\ref{sec-conclude}, we summarize our results and highlight several promising directions for future research.

%
\section{Preliminary background}\label{sec-intro}

\subsection{Double-scaled SYK and chord states\label{sec-DSSYK}}
The SYK model is a (0+1)-dimensional strongly coupled quantum mechanical system \cite{Sachdev:1992fk, Sachdev:2010um, Kitaev, Maldacena:2016hyu}. It consists of $N$ Majorana fermions with $p$-body interactions, which is described by the following Hamiltonian:
\begin{equation}\label{HAM21}
    H_{\mathrm{SYK}}=i^{p/2} \sum_{i \leq i_1 \cdots i_p \leq N} J_{i_1\cdots i_p}\psi_{i_1}\cdots \psi_{i_p}, \quad \{i,j\}=1,\cdots ,N,
\end{equation}
where the Majorana fermions follow the anti-commutation relation:
\begin{equation}
    \{ \psi_i,\psi_j\}=2\delta_{ij},
\end{equation}
and $J_{i_1\cdots i_p}$ is the random coupling constants drawn from Gaussian distributions with zero mean. The quenched averages of coupling are given by
\begin{equation}
    \brac{J_{i_1\cdots i_p}}=0, \qquad \brac{J^2_{i_1\cdots i_p}}=\frac{J^2(p-1)!}{N^{p-1}}.
\end{equation}

For convenience, we use the compact indices $I$ to denote a set of ordered $p$ indices $\{i_1,i_2,\dots, i_p\}$. In this notation, the Hamiltonian \eqref{HAM21} can be written as:
\begin{equation}
    H_{\mathrm{SYK}}=i^{p/2} \sum_{I}J_I\Psi_I.
\end{equation}
This model is recognized as being exactly solvable in the double-scaled limit\footnote{We follow notations in \cite{Blommaert:2024whf}. To match with \cite{Berkooz:2024lgq}, one can simply set $q\rightarrow\sqrt{q}$ throughout the paper.} (for a recent review, see \cite{Berkooz:2024lgq}):
\begin{equation}
    N\rightarrow \infty \,, \quad p\rightarrow \infty\,, \quad \lambda\equiv \frac{2p^2}{N} \,\, \mathrm{fixed} \,, \quad q \equiv e^{-\lambda/2} \in (0, 1) \,.
\end{equation}
In this limit, the quenched average turns to the annealed average:
\begin{equation}
    \brac{J_{I}}=0, \qquad \brac{J_I J_K}=\frac{N^2}{2p^2}\binom{N}{p}^{-1} \mathbb{J}^2\delta_{IK},
\end{equation}
where $\mathbb{J}^2 =  2|\log q|J^2$. The moments of Hamiltonian insertion can be represented by chord diagrams, which turn out to be simply \cite{Cotler:2016fpe, Berkooz:2018jqr}:
\begin{equation}
    m_n=\brac{H_{\mathrm{DSSYK}}^n}_J=\sum_{\#} q^{2\#},
\end{equation}
where the summation runs over all possible chord diagrams with $n/2$ chords and $\#$ stands for the number of intersections of chords. This progress of understanding the contraction from the chord diagrams involves the production of $n/2$-open chord states, and their closure. This brought up the concept of Hilbert space in terms of chord number state $\ket{n}$ \cite{Lin:2022rbf, Okuyama:2024yya}. The lowering and raising operators are defined as \cite{Berkooz:2024lgq}: 
\begin{equation}
    a\ket{n}=\frac{1-q^{2n}}{1-q^2}\ket{n-1}, \qquad a^\dagger \ket{n}=\ket{n+1}, \qquad \hat{n}\ket{n}=n\ket{n},
\end{equation}
where $a$ and $a^\dagger$ generate an algebra of $q$-deformed oscillator that obeys the following commutation relations:
\begin{equation}
    [a,a^\dagger]_{q^2}=aa^\dagger-q^2a^\dagger a=1, \qquad [a^\dagger,a]_{q^2}=(1-q^4)a^\dagger a-q^2,
\end{equation}
which reduces to the Heisenberg algebra of harmonic oscillators when $q \rightarrow 1$.
Another useful quantity is the transfer matrix $\hat{T}=a^\dagger+a$, which acts as a Hamiltonian in the chord state Hilbert space and generates the time evolution inside the disorder-averaged correlation function. The transfer matrix is tridiagonal in the chord basis and has a bounded spectrum: $E\in[ -2/\sqrt{1-q^2},2/\sqrt{1-q^2}]$. It is therefore convenient to
parametrize the spectrum by an angle $\theta \in [0,\pi]$ via $E(\theta)=2\cos(\theta)/\sqrt{1-q^2}$ and to define the corresponding eigenstates $|\theta\rangle$ by $\hat{T}|\theta\rangle=2\cos(\theta)/\sqrt{1-q^2}|\theta\rangle$.

When including matters, it is equivalent to considering another kind of chords in the chord diagram. Consider the matter operator as:
\begin{equation}
    M_\Delta=i^{p'/2}\sum_I J'_I \Psi'_I,
\end{equation}
where $I={i_1,i_2\dots,i_{p'}}$ and $\Delta = p'/p$. The intersection weight of this matter operator with the Hamiltonian is
\begin{equation}
    q'^2=q^{2 \Delta}.
\end{equation}
The two point function $G_{\Delta}(\beta_1,\beta_2)$ can be computed as:
\begin{equation}
    G_{\Delta}(\beta_1,\beta_2)=\brac{\Tr(e^{-\beta_1 H_{\mathrm{DSSYK}}}M_\D e^{-\beta_2 H_{\mathrm{DSSYK}}}M_\D)}_{JJ'}=\bra{0}e^{-\beta_1 T}q'^{2\hat{n}}e^{-\beta_2 T}\ket{0}.
    \label{eq-2pt DSSYK}
\end{equation}
By inserting an identity in the $\theta$-basis and chord basis, it can be written as \cite{Berkooz:2018qkz}
\begin{equation}
    \int^\pi_0 d\theta_1d\theta_2 e^{-\beta_1E(\theta_1)-\beta_2E(\theta_2)}\rho(\theta_1)\rho(\theta_2)\frac{(q^{4\D};q^2)_{\infty}}{(q^{2\D}e^{i(\pm\theta_1\pm\theta_2)};q^2)_\infty}.
\end{equation}

\subsection{Sine-dilaton gravity: a holographic dual to DSSYK}
Given a quantum theory as DSSYK, it is then natural to ask what the corresponding holographic dual would be, particularly at finite $q$ and finite temperature. A recent work \cite{Blommaert:2024ymv} proposed the sine-dilaton gravity as a candidate. Sine-dilaton gravity is one type of 2-dimensional dilaton gravity with dilaton potential given as a sine function, which is described by the following path integral:
\begin{equation}
    \int Dg D\Phi \exp\left[\frac{1}{2}\int d^2x \sqrt{g} \left( \Phi R+\frac{\sin(2|\log q |\Phi)}{|\log q|}\right)\right].
    \label{eq-SD pathInt}
\end{equation}
After rescaling the dilaton field $\Phi'=2|\log q|\Phi$, the sine-dilaton gravity action admits a classical black hole solution with cosmological horizon:
\begin{equation}
    ds^2=-F(r)dt^2+\frac{1}{F(r)}dr^2 \,, \qquad F(r) = 2(\cos \theta-\cos r) \,, \qquad \Phi'=r .
\end{equation}
With constraint of coordinate range from $r\in[0,2\pi]$, this geometry has a black hole horizon located at $r=\Phi'=\theta$ and a cosmological horizon at $r=2\pi-\theta$. In the region behind the cosmological horizon, $F(r)$ becomes negative, which suggests that light rays cannot reach out to $r<2\pi-\theta$. Expanding the metric near the asymptotic boundary, one can read off the ADM energy as
$E_{\rm ADM}=-\frac{\cos\theta}{2|\log q|}$, where $\theta$ labels the location of the
black-hole horizon \cite{Blommaert:2024ymv}. As will become clear after canonical quantization, this same parameter
labels energy eigenstates of the sine-dilaton Hamiltonian; under the DSSYK/sine-dilaton
dictionary it is identified with the $\theta$-basis introduced above. For this reason, we use
the same symbol $\theta$ on both sides and do not distinguish them in what follows. To get an AdS$_2$ black hole geometry similar to the JT gravity, one can perform a Weyl rescaling and a coordinate transformation of the metric
\begin{align}\label{METRIC216}
\begin{split}
d\tilde{s}^2&=e^{-i\Phi}ds^2=-(\rho^2-\sin^2 \theta)dt^2+\frac{1}{\rho^2-\sin^2 \theta}d\rho^2 \,,\\    \rho&=\tilde{\Phi}=i(e^{-i\Phi}+\cos\theta),
\end{split}
\end{align}
where the dilaton profile is complex with a boundary condition
\begin{equation}
    \Phi(\rho\rightarrow\infty)=\frac{\pi}{2}+i\infty.
\end{equation}
To quantize the sine-dilaton gravity, consider the Hamiltonian in the canonical formalism \cite{Harlow:2018tqv} given as
\begin{equation}
    H_{\mathrm{SD}}=\frac{1}{2|\log q|}\left( -\cos(\hat{P})+\frac{1}{2}e^{i\hat{P}}e^{-\hat{L}}\right),
    \label{eq-HamiltonianSD}
\end{equation}
where $\hat{L}$ is the geodesic length operator and $\hat{P}$ is its canonical conjugate. This Hamiltonian exactly matches the transfer matrix $\hat{T}$ in DSSYK. The holographic dual of the geodesic length operator is proposed to be proportional to the chord number operator through the following relation:
\begin{equation}
    \hat{L}=2|\log q|\hat{n}.
    \label{eq-SD DSSYK dual}
\end{equation}
This relation should be understood as a holographic identification at the operator level.
Consequently, at the disk (leading semiclassical) level, the geodesic-length eigenstate
$|L\rangle$ in sine-dilaton gravity is mapped to the chord-number eigenstate $|n\rangle$
in DSSYK. 
The wavefunction in the geodesic length basis is 
\begin{equation}
    \psi_\theta(L)=\braket{\theta}{L}=H_{L/|\log q^2|}(\cos\theta|q^2),
    \label{eq-wavefunctionSD}
\end{equation}
where $\ket{\theta}$ is the energy eigenbasis with energy $E(\theta)=-\cos\theta/|\log q^2|$  and $H$ is the q-Hermite polynomial. This matches the wavefunction in the chord number basis in DSSYK. Observing the non-Hermitian Hamiltonian in (\ref{eq-HamiltonianSD}), one consequence is that the complex conjugate of $\psi_\theta(L)$ does not equal to itself but rather
\begin{equation}
    \psi^*_\theta(L)=\braket{L}{\theta}=\frac{H_{L/|\log q^2|}(\cos\theta|q^2)}{(q^2;q^2)_{L/|\log q^2|}},
\end{equation}
where $(a;q^2)_n\equiv \prod^{n-1}_{k=0} (1-aq^{2k})$ is the q-Pochhammer symbol.

\subsection{Krylov state complexity}
In recent years, the Krylov subspace method has become increasingly important in the study of quantum chaos and black hole physics. One approach to utilizing this method is through the investigation of quantum state complexity under time evolution. 
In particular, for systems prepared in special entangled states such as the thermofield double state, this framework provides a useful tool for probing dynamical properties.
The quantity defined for this purpose is called Krylov state complexity or spread complexity \cite{Balasubramanian:2022tpr}, which has been proposed as an indicator of quantum chaos \cite{Baggioli:2024wbz}, characterized by an initial quadratic growth, followed by linear growth, and eventually reaching a peak and saturating at a finite value.

Consider a quantum system with dynamics governed by a time-independent Hamiltonian $H$ and an initial state $\ket{\varphi(0)}$. The time evolution of this state $\ket{\varphi(t)}=e^{-itH}\ket{\varphi(0)}$ naturally generates a basis $\{ \ket{\varphi_n}\}=H^n \ket{\varphi(0)}$. Applying the Gram-Schmidt procedure to this basis, we could generate an orthonormal basis, called the Krylov basis $\{\ket{K_n}\}$, where $\ket{K_0}=\ket{\varphi(0)}$. We can define the Krylov state complexity to study how much the initial state overlaps with the Krylov basis as time evolves:
\begin{equation}
    C_S(t)=\sum_n n |\braket{K_n}{\varphi(t)}|^2.
\end{equation}
The Krylov basis is the basis that minimizes this quantity in the vicinity of $t=0$. Here, one can also denote 
\begin{equation}
    p_n(t)=|\braket{K_n}{\varphi(t)}|^2,
\end{equation}
which are the probabilities of finding the time-evolved initial state in the $n$-th Krylov basis vector. The Shannon entropy of this probability distribution is the Krylov entropy or spread entropy \cite{Barbon:2019wsy}, defined as 
\begin{equation}
    S(t)=-\sum_n p_n \log p_n.
\end{equation}

Moreover, one can actually define a sequence of such complexities that are minimized by the Krylov basis. We call them higher-order Krylov complexities (or generalized spread complexity) \cite{Fu:2024fdm, Camargo:2024rrj} defined as
\begin{equation}
    C^{(m)}_S(t)=\sum_n n^m |\braket{K_n}{\varphi(t)}|^2,
\end{equation}
which are proposed to more accurately capture the pronounced peak structure for chaotic systems.

%
\section{Zoology of holographic descriptions for Krylov-space observables}\label{sec-main}
\subsection{Krylov complexity and wormhole length: a quick review}
In this subsection, we briefly review previous work \cite{Heller:2024ldz} on the duality of Krylov complexity in DSSYK and wormhole length in sine-dilaton gravity.
To start, consider that the sine-dilaton gravity action (\ref{eq-SD pathInt}) admits a Schwarzchild-like metric \eqref{METRIC216}:
\begin{equation}
    ds^2=-(\rho^2-\sin^2\theta)dt^2+\frac{d\rho^2}{\rho^2-\sin^2\theta}.
    \label{eq-AdS2 BH}
\end{equation}
This metric reduces to JT gravity at the low energy limit $\theta \ll 1$. The normalized classical wormhole length is given as
\begin{equation}
L_{\mathrm{re}}=2\log(\cosh(t\sin\theta/2))-2\log(\sin\theta)\,,
\label{eq-SD length}
\end{equation}
which is useful for the ``complexity=volume'' conjecture in two dimensions.\footnote{This CV conjecture here is strictly constrained to the 2d metric without considering the origin of JT gravity from higher-dimensional gravitational theories, where CV would be corrected by the geodesic length weighted by the dilaton profile \cite{ Bhattacharya:2023drv, Fu:2024vin, Caceres:2025myu}.} One can also obtain the geodesic length at finite temperature at the full quantum level. Consider the normalized two-point function using geodesic approximation with boundary states prepared in Euclidean signature:
\begin{equation}
    \frac{1}{Z_\beta}\bra{L=0}e^{-\tau H_{\mathrm{SD}}}e^{-\D \hat{L}}e^{-(\beta-\tau) H_{\mathrm{SD}}} \ket{L=0},
\end{equation}
where $\D$ is the scaling dimension of the boundary probe operator. This two-point function exactly matches the one from DSSYK shown in (\ref{eq-2pt DSSYK}) by setting $\beta_1=\tau$ and $\beta_2=\beta-\tau$, where $\D=p/p'$, the ratio of interacting fermions in Hamiltonian and matter.
The Lorentzian two-point function can be acquired by setting $\tau \rightarrow \beta/2+it$. To get the geodesic length, one simply takes the derivative with respect to $\D$ and sets $\D \rightarrow 0$. The explicit formula can be written in the following integral form
\begin{eqnarray}
    \brac{\hat{L}}&=&\frac{1}{Z_\beta}\int^\pi_0 d\theta_1 d\theta_2 \rho(\theta_1)\rho(\theta_2) e^{-(E(\theta_1)+E(\theta_2))\beta/2}e^{-it(E(\theta_1)-E(\theta_2))} \nonumber\\
    &&\times \sum_{L/|\log q^2|=0}^\infty \frac{L}{(q^2;q^2)_{L/|\log q^2|}}H_{L/|\log q^2|}(\cos\theta_1|q^2)H_{L/|\log q^2|}(\cos\theta_2|q^2).
\end{eqnarray}
Consider the dual in DSSYK through (\ref{eq-SD DSSYK dual}), one can translate the above formula into:
\begin{equation}
    2|\log q|\sum_n n \left\vert\bra{n} \frac{e^{-i\hat{T}(t-i\beta/2)}}{\sqrt{Z_\beta}} \ket{0}\right\vert^2,
\end{equation}
where $\ket{n}$ is the chord number state and $\hat{T}$ is the transfer matrix. The summation part of this expression has a clear physical meaning in terms of Krylov state complexity:
\begin{equation}
    C_S(t)=\sum_n n \left\vert\bra{n} \frac{e^{-i\hat{T}(t-i\beta/2)}}{\sqrt{Z_\beta}} \ket{0}\right\vert^2 \,,
\end{equation}
which measures how the zero-chord initial state located on half of the Euclidean thermal circle evolves on the chord number basis. As a result, this Krylov complexity has a clear bulk correspondence as geodesic length:
\begin{equation}
    C_S(t)=\frac{\brac{\hat{L}}}{2|\log q|}.
\end{equation}
Notably, this duality holds at the full quantum level and finite temperature. When taking the triple-scaled limit, this duality reduces to the one proposed in \cite{Rabinovici:2023yex}.

\subsection{Growth rate of Krylov complexity and wormhole velocity}
Next, given the bulk dual of Krylov state complexity, it is natural to ask what the dual of the growth rate of Krylov state complexity would be. For convenience, we denote the state $\ket{\phi(t)}=\frac{e^{-i \hat{T}(t-i\beta/2)}}{\sqrt{Z_\beta}} \ket{0}=\frac{e^{-i \hat{T} t}}{\sqrt{Z_\beta}} \ket{\phi(0)}$, which is the Euclidean and Lorentzian time-evolved zero-chord state with a normalization factor. The Krylov state complexity can be rewritten as
\begin{equation}
    C_{S}(t)=\sum_{n} \braket{\phi(t)}{n} \braket{ n}{\phi(t)} =  \bra{\phi(t)} \hat{n} \ket{\phi(t)} ,
\end{equation}
where $\sum_n\ket{n}\bra{n}\equiv \hat{n}$ is the chord number operator, which is also the Krylov operator $\hat{K}$ for the initial state being the zero-chord number state. To study the growth rate of $C_S(t)$, we consider the Heisenberg picture, in which the Krylov complexity can be written as the expectation value of the time-evolved Krylov operator with respect to the state $\ket{\phi(0)}$:
\begin{equation}
    C_{S}(t)=\bra{\phi(0)} e^{i\hat{T}t}\hat{K} e^{-i\hat{T}t} \ket{\phi(0)}.
\end{equation}
By using the Ehrenfest theorem for any time-independent operator $\hat{A}$ stated as: 
\begin{equation}
    \frac{d}{dt} \langle\hat{A}(t)\rangle=i\langle[H,\hat{A} ]\rangle \,.
\end{equation}
We find the growth rate of Krylov state complexity as
\begin{eqnarray}
    \frac{d}{dt}C_{S}(t)&=& \bra{\phi(t)}i[\hat{T},\hat{n}] \ket{\phi(t)}\nonumber\\
    &=& \frac{1}{Z_\beta}\int^\pi_0 d\theta_1 d\theta_2 \rho(\theta_1)\rho(\theta_2) i (E(\theta_1)-E(\theta_2)) e^{-(E(\theta_1)+E(\theta_2))\beta/2}e^{-it(E(\theta_1)-E(\theta_2))}\nonumber\\
    &&\times\sum_n^{\infty}\frac{n}{(q^2;q^2)_n} H_n(\cos\theta_1\vert q^2)H_n(\cos\theta_2\vert q^2) \,.
    \label{eq-Cs growth rate}
\end{eqnarray}
On the sine-dilaton gravity side, we find that a similar object is the wormhole velocity operator \cite{Iliesiu:2024cnh}, which is defined as a quantity conjugate to the wormhole length:
\begin{equation}
    \hat{\pi}=i[H_{\text{SD}},\hat{L}].
\end{equation}
The expectation value of this operator can be written as
\begin{eqnarray}
    \brac{\hat{\pi}}&=&i\bra{L=0} e^{-\tau H_{sg}}[H_{\text{SD}},\hat{L}] e^{-(\beta-\tau)H_{\text{SD}}}  \ket{L=0}\nonumber\\
    &=&-i\frac{\partial}{\partial\Delta} \left.\left[ \bra{L=0} e^{-\tau H_{\text{SD}}}[H_{\text{SD}},e^{-\Delta \hat{L}}] e^{-(\beta-\tau)H_{\text{SD}}} \ket{L=0}\right]\right\vert_{\Delta \rightarrow 0}.
    \label{eq-wornhole velocity}
\end{eqnarray}
We can then do a Wick's rotation: $\tau=\beta/2+it$ to acquire the Lorentzian expectation value of wormhole velocity.

By applying the holographic dictionary (\ref{eq-SD DSSYK dual}), we can immediately find the duality of the growth rate of Krylov complexity in DSSYK and the wormhole velocity in sine-dilaton gravity:
\begin{equation}
    \frac{d}{dt}C_{S}(t)=\frac{\brac{\hat{\pi}(t)}}{2|\log q|}.
\end{equation}
This duality is straightforward to observe through simple quantum mechanical commutation relations, as done above. Nevertheless, it is more interesting to discuss the physical interpretation of this duality. Namely, we would like to study in depth the growth rate of Krylov complexity in DSSYK and how this quantity can probe the physics in the bulk.
For this purpose, let us consider the wormhole velocity operator in DSSYK given as:
\begin{equation}
    \hat{\pi}_{D}=i[\hat{T},\hat{n}].
\end{equation}
From Section \ref{sec-DSSYK}, we know that the transfer matrix $\hat{T}$ in DSSYK is written as 
\begin{equation}
    \hat{T}=a^\dagger+a, \hspace{0.8cm} a=\alpha \frac{1-q^{2\hat{n}}}{1-q^2}, \hspace{0.8cm} a^\dagger=\alpha^\dagger,
\end{equation}
where $\alpha \ket{n}=\ket{n-1}$ and $\alpha^\dagger\ket{n}=\ket{n+1}$. Since $[a^\dagger,\hat{n}]=-a^\dagger$, $[a, \hat{n}]=a$, the DSSYK wormhole velocity operator is thus
\begin{equation}
    \hat{\pi}_{D}=i(a-a^\dagger).
    \label{eq-wormholeVDSSYK}
\end{equation}
Therefore, the operator $\hat{\pi}_{D}$ acting on chord states would create or annihilate a chord, which seems to correspond to the expanding or shrinking (firewall) of the wormhole in the dual picture \cite{Iliesiu:2024cnh,Stanford:2022fdt}. Operator $\hat{\pi}_{D}$ has a nice Toeplitz tri-diagonal matrix form in the chord number basis:
\begin{equation}
\brac{\hat{\pi}_{D}}=i
\begin{pmatrix}
0 & -1 & &&\\
\frac{1-q^2}{1-q^2} & 0 & -1&&\\
 & \frac{1-q^4}{1-q^2} & \ddots&\ddots&\\
 &  & \ddots&\ddots&-1\\
 &  & &\frac{1-q^{2(N-1)}}{1-q^2}&0\\
\end{pmatrix},
\end{equation}
where $N$ is the dimension of the chord Hilbert space. Following \cite{Berkooz:2018qkz}, similar to the transfer matrix, we can see that this operator has a bounded spectrum in the large-$N$ limit:
\begin{equation}
    \left(-\frac{2}{\sqrt{1-q^2}},\frac{2}{\sqrt{1-q^2}} \right).
\end{equation}
Although this spectrum is identical to the transfer matrix, it is worth noticing that they do not share the same eigenstates. Thus, the commutator of $T$ and $\hat{\pi}_{D}$ is nonzero. Given the form of the wormhole velocity operator (\ref{eq-wormholeVDSSYK}) in DSSYK, it is natural to consider its expectation values in coherent states. This is because the coherent state intrinsically diagonalizes the annihilation operator, which leads to expectation values of operators like the wormhole operator behaving simply and classically. As we will see in the next subsection, this turns out to be highly useful for interpreting the operator in terms of wormhole dynamics and firewalls.

\subsubsection{Coherent state in DSSYK and bulk description \label{sec-Coherent DSSYK}}
Knowing from Section \ref{sec-DSSYK} that $a$ and $a^\dagger$ in DSSYK form the algebra of $q$-deformed oscillator:
\begin{equation}
    a \ket{n}=[n]_{q^2} \ket{n-1}, \qquad a^\dagger\ket{n}=\ket{n+1},
    \label{eq-q-deformed algebra}
\end{equation} 
where $[n]_{q^2}\equiv \frac{1-q^{2n}}{1-q^2}$ is the $q$-numbers, it is natural to consider the coherent state $\ket{z}$. To find the explicit form of it, we consider expanding it in the chord state basis:
\begin{equation}
    \ket{z}=\sum^\infty_{n=0} c_n \ket{n},
    \label{eq-coherent expansion}
\end{equation}
where $c_n=\braket{z}{n}$ are overlaps of coherent state and the chord number state. One of the definitions of a coherent state is that it is a unique eigenstate of the annihilation operator\footnote{Note that there are three different definitions of coherent state. Except for the first one we mentioned here, the second and third ones are group theoretical definitions through the displacement operators and definitions through minimal uncertainty. They generally may not be equivalent to each other in certain scenarios \cite{Perelomov:1986tf} (in fact, they only agree in the case of the Heisenberg-Weyl group). However, in our case, we showed the equivalent of the first and second definitions in (\ref{eq-coherent state Dis}) in $q$-deformed oscillators.}:
\begin{equation}
    a\ket{z}=\sum^\infty_{n=0} c_n z\ket{n}.
    \label{eq-coherent1}
\end{equation}
Plug (\ref{eq-coherent expansion}) to this definition together with (\ref{eq-q-deformed algebra}), we have
\begin{equation}
    a\ket{z}=\sum^\infty_{n=0} c_{n+1}[n+1]_{q^2} \ket{n}.
    \label{eq-coherent2}
\end{equation}
By comparing (\ref{eq-coherent1}) and (\ref{eq-coherent2}), we can find the recursion relation for coefficient $c_n$:
\begin{equation}
    c_{n+1}[n+1]_{q^2}=zc_n.
\end{equation}
We can therefore conclude:
\begin{equation}
    c_n=\frac{z^n}{[n]_{q^2}!}c_0,
\end{equation}
where $[n]_{q^2}!\equiv \prod^{n}_{k=1} [k]_{q^2}$ is the $q$-factorial. Eventually, we find the normalized coherent state:
\begin{equation}
    \ket{z}=\frac{1}{\sqrt{e^2_{q^2}(|z|)}}\sum^\infty_{n=0}\frac{z^n}{[n]_{q^2}!}\ket{n},
\end{equation}
where $|c_0|^2=\frac{1}{e^2_{q^2}(|z|)}$ by imposing $\braket{z}{z}=1$, and we denote $e^2_{q^2}(|z|)$ as the q-deformed squared exponential defined as:
\begin{equation}
e^2_{q^2}(|z|^2)\equiv \sum^\infty_{n=0}\frac{|z|^{2n}}
{([n]_{q^2}!)^2} .
\end{equation}
\textit{This is the normalized coherent state for the q-deformed oscillator} \cite{eremin2006q}.
It is then straightforward to observe that this coherent state can be obtained by acting the q-deformed displacement operator on the zero-chord state:
\begin{equation}
    \ket{z}=D_{q^2}(z)\ket{0},
    \label{eq-coherent state Dis}
\end{equation}
where
\begin{equation}
    D_{q^2}(z)=\frac{e_{q^2}(za^\dagger) e_{q^2}(\bar{z}a)}{e^2_{q^2}(|z|^2}),
\end{equation}
here $e_q(x)=\sum^\infty_{n=0}\frac{x^n}{[n]_{q^2}!}$ is the q-deformed exponential (or q-exponential). Before moving on,  we would like to comment on the chord basis in (\ref{eq-q-deformed algebra}). This widely used notation for the chord basis $\{\ket{n}\}$ is perfectly fine for DSSYK computations. However, this chord basis is known to be the orthogonal basis with the inner product:
\begin{equation}
\braket{n}{m}=[n]_{q^2}!\delta_{nm}.
\end{equation}
Therefore, it is imperative to note that this usual chord basis cannot play the role of a Krylov basis, which is defined to be orthonormal through the Lanczos algorithm. The correct basis for the Krylov space should be the orthonormal chord basis $\{\ket{\tilde{n}}\}$:
\begin{equation}
    a \ket{\tilde{n}}=\sqrt{[n]_{q^2}} \ket{\widetilde{n-1}}, \qquad a^\dagger\ket{\tilde{n}}=\sqrt{[n+1]_{q^2}}\ket{\widetilde{n+1}},
\end{equation}
which obeys the orthonormality condition:
\begin{equation}
    \braket{\tilde{n}}{\tilde{m}}=\delta_{\tilde{n}\tilde{m}}.
\end{equation}
Nevertheless, this change of chord basis does not affect the result of complexity if one does it properly with normalization. In this orthonormal chord basis, the displacement operator for the coherent state will be changed to:
\begin{equation}
    \tilde{D}_{q^2}(z)=\frac{e_{q^2}(za^\dagger) e_{q^2}(\bar{z}a)}{e_{q^2}(|z|^2)},
\end{equation}
where only the normalization factor in the denominator is just the $q$-exponential.
This coherent state can also be prepared through the Euclidean path integral \cite{Skenderis:2008dg, Botta-Cantcheff:2015sav, Marolf:2017kvq}\footnote{Another way of preparing a semi-classical state is to slice the Euclidean wormholes through a time-reflection symmetric slice, which yields an order one overlap with the TFD state semi-classically \cite{Hubeny:2007xt, Belin:2025wju}.}, which can be done via an inserted source:
\begin{equation}
    \ket{z}=Te_{q^2}^{-\int^0_{-\infty} d\tau J_\alpha(\tau)\Psi_\alpha(\tau)} \ket{0},
\end{equation}
where $T$ means time ordering, $e_{q^2}$ is the q-deformed exponential, $J_\a(\tau)$ is the source. This is a general Euclidean state preparation procedure for any operator $\Psi_\a$ and vacuum state $|0\rangle$ that belong to the same Hilbert space. Since we are working in the H-chord basis, $J_\a \Psi_\alpha$ can be taken as the Hamiltonian H (or any composite operators well defined in the chord Hilbert space that can be decomposed as modes of $a$ and $a^\dagger$). Thus, to prepare the coherent state in the H-chord Hilbert space, one shall take:
\begin{equation}
    \ket{z}=Te_{q^2}^{-\int^0_{-\infty} d\tau z(\tau)H_{\mathrm{DSSYK}}(\tau)} \ket{0},
\end{equation}
where this perturbation is the large-$p$ generalization of the ones introduced in \cite{Garcia-Garcia:2020ttf, Arias:2023ygy}. Certainly, $H$ contains both creation and annihilation operators in the chord basis. However, when acting on the zero-chord state $\ket{0}$, only the creation operators contribute. The mixed terms proportional to the number operator contribute only a constant. Therefore, the Euclidean path integral prepared state can be written as
\begin{equation}
    \ket{z}\propto e_{q^2}^{z a^\dagger}\ket{0},
    \label{eq-Euclidean Path Int}
\end{equation}
where $z$ contains the information of the source and plays the role of a coherent parameter. This matches the unnormalized coherent state we found in (\ref{eq-coherent state Dis}).

We then move on to evaluate the wormhole velocity operator in the coherent state in DSSYK:
\begin{equation}
\bra{z} i(a^\dagger-a) \ket{z}=2\Im(z).
\label{eq-wormhole velocity DSSYK}
\end{equation}
Given the coherent state as a superposition of chord states with weights, its bulk dual can be interpreted as a collection of geodesic paths (a wavepacket on the geodesic basis), which can be probed by classical bulk fields. Next, we will simply show that $\Im(z)$ corresponds to a time shift of field excitations in the bulk description. Consider the boundary one-point function in the coherent basis:
\begin{equation}
    \bra{z}\mathcal{O}(t) \ket{z},
\end{equation}
which is finite in the coherent basis. Here, operator $\hat{O}$ is a composite operator built from $a$ and $a^\dagger$ in the $q$-deformed oscillator, whose one-point function can still be non-trivial.\footnote{One should not consider this as a matter operator, whose one-point function certainly vanishes. We thank Sergio E. Aguilar-Gutierrez for pointing this out.} Acting explicitly on the chord basis, it is given by
\begin{equation}
   \bra{z}\mathcal{O}(t) \ket{z}=N^2(|z|) \sum_n c^2_n\bra{n}\mathcal{O}(t) \ket{n}+\sum_{m \neq n} c_m^* c_n\bra{m}\mathcal{O}(t) \ket{n},
\end{equation}
where $N(|z|)$ is the normalization factor, $c_n=z^n/[n]_{q^2}!$, and
\begin{equation}
    c_m^* c_n=\frac{|z|^{m+n}e^{i(n-m)\arg(z)}}{[n]_{q^2}![m]_{q^2}!}.
\end{equation}
Using the HKLL reconstruction map \cite{Hamilton:2006az}, the bulk field can be reconstructed by
\begin{equation}
    \brac{\phi(r,t)}=\int dt' K(r,t;t')\bra{z}\mathcal{O}(t') \ket{z},
\end{equation}
where $\langle \rangle$ denotes the expectation value under the bulk corresponding state of $\ket{z}$. Taking the relevant part of the smearing function $K(r,t;t')$\footnote{Note that although HKLL reconstruction was proposed in AdS, where the smearing function can have a nice form, which may not be true in dS. However, in our arguments, we only care about the Fourier modes, which shall hold generally.}, it can be expressed as:
\begin{equation}
    \brac{\phi(r,t)}\sim \sum_{n,m}\sum_\omega e^{-i\omega (t+(n-m)\arg(z)/\omega)} f_\omega(r)+\mathrm{c.c.}~, 
    \label{eq-bulk mode shift}
\end{equation}
which indicates a time shift of the corresponding bulk field modes. Take the first term as an example. Consider $n>m$ and modes with $\vert(n-m)\vert/\omega\approx 1$. Using the definition of argument of a complex number: $\arg(z)=\arctan(\Im(z)/\Re(z))\in (-\pi,\pi]$ when $\Re(z)>0$, a negative imaginary part implies $\arg(z)\in (-\pi,0)$. This corresponds to a backward shift in time for the bulk field modes, as illustrated in the figure \ref{Fig-Firewall} (a). Likewise, for modes with $n<m$, the positive imaginary part of the coherent parameter would contribute to the backward shift in time. Effectively, this results in a relative boost of the left perturbation depicted in the figure \ref{Fig-Firewall} (b), such that a collision of the infalling probe and a perturbation can happen. This matches the interpretation of black hole/white hole tunneling. As shown in the figure \ref{Fig-Firewall} (b), the orange dashed line represents an infalling probe with no boost, which does not intersect with the solid green curve. However, after having a backward shift in time, the probe (dashed orange line) collides with the dashed green curve, effectively producing a firewall. This scenario corresponds to the Stanford-Yang \cite{Stanford:2022fdt} type of firewall (see also \cite{Balasubramanian:2024lqk} for comments related to Krylov complexity and firewalls). To summarize, in DSSYK, the wormhole velocity operator captures the imaginary part of the coherent parameter, which, through bulk reconstruction, indicates the possibility of experiencing a firewall.

\begin{figure}
\hspace{1cm}
\begin{tikzpicture}
\draw(-4,0)--(-4,4);\draw(0,0)--(0,4);
\draw(-4,0)--(0,4);\draw(-4,4)--(0,4);
\draw(-4,4)--(0,0);\draw(-4,0)--(0,0);
\draw[blue] (0,0) arc [radius=2.6, start angle=230, end angle=130];
\draw[red](-4,2)--(0,2);
\draw [blue]   (-1,1) to[out=135,in=-90] (-1.5,2);
\draw[blue](-1,1)--(0,0);
\draw[blue]    (-1.5,2) to[out=90,in=225] (-1,3);
\draw[blue](-1,3)--(0,4);
\draw[fill=orange,orange](-1.5,2) circle [radius=0.05];
\draw[->][thick](-1.5,1.9)--(-1.5,1.6);
\draw[fill=orange,orange](-0.93,2) circle [radius=0.05];
\draw[->][thick](-0.93,1.9)--(-0.93,1.6);
\node at (-0.7,1.7){$\phi$};
\node at (-2,-0.5){$(a)$};
\draw[green,thick] (-2,0)--(-4,2);
\draw[green,thick] (-4,2)--(-2,4);

\draw[->](1,2)--(3,2);
\node at (2,2.5){\small{effectively}};

\draw(4,0)--(4,4);\draw(8,0)--(8,4);
\draw(4,0)--(8,4);\draw(4,4)--(8,4);
\draw(4,4)--(8,0);\draw(4,0)--(8,0);
\draw[orange,dashed,thick](8,2.2)--(6.2,4);
\node [right]at(8,2) {\small{infalling probe}};
\node at (6,-0.5){$(b)$};
\draw[green,thick] (4,2)--(6,4);
\draw[->][thick](5,2.9)--(5,2.6);
\draw[green,thick,dashed] (4,1.5)--(6.5,4);
\node[star, star points=8, draw, minimum size=0.3cm,inner sep=0pt,fill=blue,star point ratio=2.25] at (6.4,3.8) {};
\end{tikzpicture}
\caption{(a) shows the backward shift in time of modes for bulk field $\phi$ according to equation (\ref{eq-bulk mode shift}). The blue curves are constant $r$ slices, and the red line is a constant time slice. (b) shows an effective picture with bulk modes of the infalling probe unchanged, but a relative boost of the left perturbation. The dashed green curve represents a boosted perturbation. The orange dashed line stands for an unchanged infalling probe. The blue star represents a collision of the infalling probe and the perturbation.\protect\footnotemark}
\label{Fig-Firewall}
\end{figure}
\footnotetext{We thank Zhenbin Yang for pointing out a misleading part in the previous plot.}

Moreover, we would like to comment on several notable aspects regarding quantum information complexities with coherent states. Given the coherent state, it is straightforward to study its quantum information geometry, i.e., the Fubini-Study metric. To proceed, we start with the expression for the coherent state and observe that
\begin{equation}
    \partial_{\bar{z}} \ket{z}=0, \qquad \partial_z \bra{z}=0.
\end{equation}
We define the K\"ahler potential for the unnormalized coherent state:
\begin{equation}
    K(z,\bar{z})=\log(\braket{z}{z})=\log(e^2_{q^2}(|z|)).
\end{equation}
The Fubini-Study metric is defined as:
\begin{equation}
    ds^2=\partial_z \partial_{\bar{z}} K(z,\bar{z}) dzd\bar{z}\equiv g_{z\bar{z}}dzd\bar{z},
    \label{eq-Fubini-study}
\end{equation}
where $z, \bar{z}$ play the roles of coordinates in a complex manifold with the K\"ahler metric $g_{z\bar{z}}$. Therefore, there exists a closed K\"ahler form, which is also a symplectic form, written as
\begin{equation}
    \Omega=i\partial_z \partial_{\bar{z}} K(z,\bar{z}) dz \wedge d\bar{z}.
\end{equation}
In the language of the Euclidean path integral, as in (\ref{eq-Euclidean Path Int}), this symplectic form is prepared at the boundary of the Euclidean manifold. It can be pushed into the bulk to any Cauchy surfaces anchored at $t_E=0$ through the ``new York" deformation \cite{Belin:2018fxe, Belin:2018bpg, Pedraza:2021mkh,Pedraza:2021fgp}. Using the piece-wise holographic dictionary \cite{Skenderis:2008dh, Skenderis:2008dg}, one can correspond the boundary source $z$ to the bulk classical field configuration $\phi$:\footnote{The inverse problem, namely ``bulk reconstruction from complexity,'' yields a precise notion of emergence of spacetime from optimized computation in the dual CFT~\cite{Pedraza:2022dqi,Carrasco:2023fcj}.}
\begin{equation}
    \phi(t,x)=\phi^+(t,x)+\phi^-(t,x),
\end{equation}
and
\begin{equation}
    \phi^\pm(t,x)=\int_{\mathcal{\partial M}^\pm} G_E(t,x;t) z^\pm(t),
    \label{eq-source-bulk field}
\end{equation}
where $z^\pm(t)$ is defined as 
\begin{equation}
    z^\pm(t)=z(t), ~~~\mathrm{for}~~t\in \mathcal{\partial M}^+; \qquad  z^\pm(t)=\bar{z}(t), ~~~\mathrm{for}~~t\in \mathcal{\partial M}^-.
\end{equation}
Consider any initial Cauchy slice $\Sigma_0$, the initial position $\varphi$ and conjugate momentum $\pi$ are related to the bulk field configuration on $\Sigma_0$:
\begin{equation}
    \varphi(t_0,x_0)=\mathrm{Re}[\phi^+(t_0,x_0)], \qquad \pi(t_0,x_0)=\mathrm{Im}[\partial_n\phi^+(t_0,x_0)],
\end{equation}
where $\partial_n$ denotes the derivative in the normal direction of $\Sigma_0$. Consider (\ref{eq-source-bulk field}) and the fact that $z^+$ and $z^-$ are complex conjugate to each other; the initial position and momentum are thus real valued. Especially, the momentum is proportional to the imaginary part of the source:
\begin{equation}
    \pi(t_0,x_0) \propto \mathrm{Im}[z(t_0)],
\end{equation}
which surprisingly matches with our result in (\ref{eq-wormhole velocity DSSYK}).

It would also be interesting to compute the volume associated with this Fubini-Study metric (\ref{eq-Fubini-study}) and compare it with the complexity of the initial coherent state, despite the absence of an analytical expression for the metric. Additionally, it would be worthwhile to study the Fubini-Study cost function proposed in \cite{Erdmenger:2022lov}. However, these directions lie beyond the scope of the present work and are left for future investigation.

\subsubsection{Bulk proper momentum}
So far, we have studied the duality between the growth rate of Krylov complexity and the wormhole velocity. However, a recent work \cite{Caputa:2024sux} proposed an alternative duality between the Krylov complexity rate and the proper momentum of a massive particle falling into the bulk geometry.\footnote{See also \cite{Fan:2024iop,He:2024pox} for the massless case.} Motivated by this, we investigate such duality and its limitations. Starting from the AdS$_2$ black hole geometry given in (\ref{eq-AdS2 BH}), we perform the following coordinate transformation to obtain a flat-space coordinate that measures the proper distance to the black hole horizon:
\begin{equation}
    \rho=\sin\theta \cosh r.
\end{equation}
The metric turns into
\begin{equation}
    ds^2=-\sin^2\theta \sinh^2 r dt^2+dr^2.
\end{equation}
This geometry is similar to the case studied in \cite{Caputa:2024sux}. Consider an infalling massive particle following a time-like trajectory:
\begin{equation}
    \tanh r(t)=\frac{ \tanh r(0)}{\cosh(t\sin\theta )}.
\end{equation}

After rescaling, the proper momentum in $r$-direction is then given as
\begin{equation}
    P_r\propto \sinh\left(\frac{\sin\theta}{2} t\right).
    \label{eq-Proper momentum}
\end{equation}
Compared to the rate of classical geodesic length (\ref{eq-SD length}):
\begin{equation}
    \frac{d L_{\mathrm{re}}}{dt}=\sin \theta  \tanh \left(\frac{\sin\theta}{2} t  \right),
    \label{eq-rate of Geo}
\end{equation}
 two results do not match exactly. However, keep in mind that this classical geodesic length is accurate only in early time, where we can assume $\cosh\left(\frac{\sin\theta}{2} t\right) \approx 1$. By adding the cosh term to the proper momentum, we find that
 \begin{equation}
     P_r\propto \tanh\left(\frac{\sin\theta}{2} t\right),
     \label{eq-small t}
 \end{equation}
which matches well with the growth rate of classical geodesic length in the early time. 
Rather, one can also consider the low-energy limit where $\theta \ll 1$, (\ref{eq-Proper momentum}) and (\ref{eq-rate of Geo}) also match in this limit:
\begin{equation}
    \frac{d L_{\mathrm{re}}}{dt} \propto P_r=\frac{\theta}{2}t,
    \label{eq-small theta}
\end{equation}
where they both show a linear dependence in time. From equations (\ref{eq-small t}) and (\ref{eq-small theta}), we see that the growth rate of Krylov complexity and proper momentum agree with each other at both low energy and early time limits. This is not an accidental consequence. In the low-energy limit, the cosmological horizon disappears and sine-dilaton gravity reduces to JT gravity. For the proper momentum of a particle inserted at $t=0$, it probes the geometry only near the boundary at early times, where the geometry of sine-dilaton gravity agrees with JT gravity asymptotically. Thus, these two limits essentially probe the same effective geometry, which yields the agreement. 

It is worth comparing our results with Ref.~\cite{Caputa:2024sux}, where an exact correspondence between
the proper radial momentum and the growth rate of Krylov (spread/state) complexity is
found to hold at all times. In contrast, we find that this duality does not hold exactly in
our setting. This should not be viewed as a counterexample to Ref.~\cite{Caputa:2024sux}. Rather, we
study a more general (and more microscopic) setup in which the momentum-complexity
relation holds only in the Schwarzian limit, i.e.\ at low energies/early times. From the boundary perspective, Ref.~\cite{Caputa:2024sux} considers a situation in which the relevant dynamics closes on an $\mathrm{SL}(2,\mathbb{R})$ algebra. By contrast, in DSSYK we probe the full spectrum, and the dynamics is governed by $\mathrm{SL}(2,\mathbb{R})$ only in the Schwarzian regime. Our results therefore suggest that the duality between proper momentum and the growth rate of complexity can be sensitive to UV physics and may receive nonperturbative corrections. From the bulk perspective, this is also natural: since DSSYK is a disorder-averaged (ensemble-type) theory, its gravitational dual is expected to exhibit strong nonperturbative effects beyond the universal IR. These effects become important at later times/higher energies and can account for the observed deviations from the probe-momentum picture away from the Schwarzian limit.

From the above observations, we argue that the statement presented in \cite{Caputa:2024sux} appears to hold only at the early times or the low energy limit of our setting. The dual of the growth rate of Krylov state complexity can be summarized as follows: 
\begin{equation}
   \mathrm{Wormhole~~ velocity} \quad \xleftrightarrow{\mathrm{~~~~~exact~~~~~}} \quad  \mathrm{Rate ~~of~~}C_S(t) \quad \xleftrightarrow[\mathrm{low~~ energy}]{\mathrm{early ~~time/} } \quad \mathrm{Proper~~ momentum}.
\end{equation}

Beyond the Schwarzian (early-time/low-energy) limit, where the leading term in the growth
rate of Krylov state complexity is dual to the proper radial momentum, it would be very
interesting to understand whether the subleading terms admit a direct gravitational
interpretation. A natural expectation is that they correspond to corrections beyond the
leading semiclassical approximation, e.g.\ finite-$G_N$ effects (and more generally
corrections to the JT/Schwarzian approximation in sine-dilaton gravity), possibly including
ensemble/non-perturbative effects. Since a systematic analysis of these corrections is beyond
the scope of this paper, we leave it for future work.

\subsection{Higher-order Krylov complexity and replica wormholes}
In DSSYK, higher-order Krylov complexity can be generated by the generating function \cite{Qi:2018bje, Xu:2024gfm}:
\begin{equation}
    I_\mu(t)=\bra{\phi(t)}e^{-\mu \hat{n}}\ket{\phi(t)},
\end{equation}
which contains information for all moments of $\hat{n}$ evaluated in the state $\ket{\phi(t)}$. To extract each moment, we take derivatives of $I_\mu(t)$ to get Krylov complexity at any order $m$:
\begin{equation}
    C^{(m)}(t)=(-1)^m\left.\frac{\partial^m}{\partial\mu^m}I_\mu(t)\right\vert_{\mu\rightarrow0}.
\end{equation}
The higher-order Krylov complexities are thus written as
\begin{equation}
    C^{(m)}(t)=\bra{\phi(t)} \hat{n}^m \ket{\phi(t)}=\frac{\Tr_\beta(U^\dagger\hat{n}^mU)}{Z_\beta},
\end{equation}
where $Z_\beta=\Tr(e^{-\beta H})$, $U$ is the Lorentzian time evolution, and $\Tr$ is the thermal trace taken in the zero-chord state. 
This quantity is closely related to a quantity we call \textit{logarithmic Krylov complexity}, denoted as $LC$ \footnote{This is different from $\log(C(t))$. More detailed analysis of logarithmic Krylov operator complexity in quantum systems is shown in \cite{Camargo:2026szl}.}:
\begin{equation}
    LC(t)=\bra{\phi(t)} \log\hat{n} \ket{\phi(t)} \,.
\end{equation}
This can be computed using the replica trick:
\begin{equation}
    LC(t)=\lim_{m\rightarrow 0} \frac{d C^{(m)}(t)}{dm}=\lim_{m\rightarrow 0}\frac{C^{(m)}(t)-1}{m}  \,.
\end{equation}
When $m$ is only taken as a positive integer, $C^{(m)}(t)$ reduces to the higher-order complexity discussed in \cite{Fu:2024fdm, Camargo:2024rrj}. Nevertheless, $m$ is, in general, a continuous parameter that requires analytical continuation.
Notice that this replica trick is different from the one used for entanglement entropy, but rather close to the one shown up in the free energy calculations \cite{Engelhardt:2020qpv}. To be more precise, the higher-order Krylov complexities should be written in terms of the ensemble-averaged quantities:
\begin{equation}
    C^{(m)}(t)=\frac{\Tr_\beta(\hat{n}(t)^m)}{Z_\beta} \,,
\end{equation}
where it already represents the ensemble average of samples of random Hamiltonians in the chord Hilbert space. Note that $\hat{n}(t)$ can be understood as prepared from $\t=0$ by a Euclidean path integral in imaginary time. Unlike in \cite{Calabrese:2005in}, the manifolds at $\t=0$ are cut open since $\hat{n}(t)$ has support on the entire geometry. They are then connected cyclically on replicated manifolds, see figure \ref{fig-replicas}(a) as an example for $m=3$. It is worth noting that the crosses on the plot do not mean the edge of a subregion, as in entanglement entropy. They are simply notations of matrix indices of $\hat{n}(t)$. In this case, one cannot regard the Krylov operator (chord number operator) as a density matrix because its trace is not bounded. Instead, one can break down the summation and deal with each term separately:
\begin{equation}
    \Tr_\beta(\hat{n}(t)^m)=\Tr_\beta(\hat{n_1}(t)^m)+2^m \Tr_\beta(\hat{n_2}(t)^m)+\dots,
\end{equation}
where $\hat{n}_1=\ket{1}\bra{1}$,  $\hat{n}_2=\ket{2}\bra{2}$, and so on. In this way, one can regard it as a reduced density matrix in each term, with the traces as 1.
\begin{figure}[h!]
    \centering
\begin{tabular}{ccc}
       \hspace{-1cm}\begin{tikzpicture}[line/.style={-},thick,scale=0.7]

\foreach \y in {-0.5, -2, -3.5} {
    \draw[thick] (0,\y) ellipse (1.8 and 0.6);
}
\node[]  at (-0.6,-0.5) {$\bigtimes$};
\node[]  at (0.6,-0.5) {$\bigtimes$};
\node[] at (-0.6,-2) {$\bigtimes$};
\node[]  at (0.6,-2) {$\bigtimes$};
\node[]  at (-0.6,-3.5) {$\bigtimes$};
\node[]  at (0.6,-3.5) {$\bigtimes$};

 \path [red] (0.6,-3.5) edge (-0.6,-3.5);
  \path [red] (-0.6,-2) edge (0.6,-2);
  \path [red] (-0.6,-0.5) edge (0.6,-0.5);
    \path [bend left=70, red] (-0.6,-3.5) edge (-0.6,-2);
    \path [bend right=70, red] (0.6,-2) edge (0.6,-0.5);
    \path [bend right=55, looseness=1.4, red] (-0.6,-0.5) edge (-3,-2);
    \path [bend right=70, looseness=1.4, red](-3,-2) edge (0.6,-3.5);
\end{tikzpicture}
 &
        \begin{tikzpicture}[line/.style={-},thick, scale=0.6]

    \draw[thick,rotate=30] (0,0) ellipse (1.5 and 0.3);
    \draw[thick,rotate=-30] (4,2.1) ellipse (1.5 and 0.3);
    \draw[thick] (2,-3) ellipse (1.5 and 0.3);
\path [bend right=80, red] (-1.2,-0.8) edge (1.3,0.8);
\path [bend right=80, red] (3.2,0.6) edge (5.7,-1);
\path [bend left=80, red] (0.5,-3) edge (3.5,-3);
\end{tikzpicture} & 
 \begin{tikzpicture}[line/.style={-},thick, scale=0.6]
 \draw[thick,rotate=30] (0,0) ellipse (1.5 and 0.3);
    \draw[thick,rotate=-30] (4,2.1) ellipse (1.5 and 0.3);
    \draw[thick] (2,-3) ellipse (1.5 and 0.3);
\path [bend left=80, red] (-1.2,-0.8) edge (0.5,-3);
\path [bend right=80, red] (1.3,0.8) edge (3.2,0.6);
\path [bend right=80, red] (5.7,-1) edge (3.5,-3);
\end{tikzpicture}        \\
       (a)& (b) & (c)  \\
    \end{tabular}
    \caption{(a) The schematic picture of replicated manifolds with three replicas in DSSYK. The crosses label the indices for the matrix $\hat{n}_{ij}(t)=(U^\dagger\hat{n}U)_{ij}$, which is no longer a diagonal matrix in general. The red lines indicate the contraction of indices.  (b) The schematic picture of the disconnected saddle in the bulk gravity. (c) The schematic picture of the connected replica wormhole saddle in the bulk gravity.}
    \label{fig-replicas}
\end{figure}

On the bulk side, using the holographic correspondence (\ref{eq-SD DSSYK dual}), the higher-order complexity can be computed through the following expressions:
\begin{equation}
    C^{(m)}(t)\longleftrightarrow \int dL \bra{L_0}\left(L\ket{L(t)}\bra{L(t)}\right)^m\ket{L_0}=-\left.\frac{\partial}{\partial \D}\Tr\left( e^{-\D \hat{L}^m(t)}  \right)\right\vert_{\D \rightarrow 0},
\end{equation}
where the integral is the Euclidean gravitation path integral. Or, from another perspective, one can also obtain it via:
\begin{equation}
    C^{(m)}(t)\longleftrightarrow \int dL \bra{L_0}\left(L\ket{L(t)}\bra{L(t)}\right)^m\ket{L_0}=(-1)^m\left.\frac{\partial^m}{\partial \D^m}\Tr\left( e^{-\D \hat{L}(t)}  \right)\right\vert_{\D \rightarrow 0}.
\end{equation}
Our interpretations of these two perspectives are: the first one can be understood as a particle with scaling dimension $\Delta$ sent to a replicated bulk geometry to probe the replicated geodesic, which would include contributions through the replica wormholes.  The second perspective can be viewed as sending many light probe particles to probe a fixed bulk geometry. Figure \ref{fig-replicas}(b)  and \ref{fig-replicas}(c) show examples of connected and disconnected geometry in the case of three replicas. We believe these two perspectives shall be equivalent at early times when replica wormhole contributions can be neglected. However, it was shown in \cite{Engelhardt:2020qpv} that connected topologies might dominate in low temperatures, where these two perspectives shall deviate from each other dramatically.

 Similarly, by using the replica trick, the log Krylov complexity can also be computed in the bulk from two perspectives: 
\begin{eqnarray}
    LC(t) &\longleftrightarrow& -\left.\frac{\partial}{\partial m}\left.\frac{\partial}{\partial\Delta}\Tr\left( e^{-\D \hat{L}^m(t)}  \right)\right\vert_{\Delta\rightarrow0}\right\vert_{m\rightarrow0},\nonumber\\
    &\longleftrightarrow& (-1)^m\left.\frac{\partial}{\partial m}\left.\frac{\partial^m}{\partial \D^m}\Tr\left( e^{-\D \hat{L}(t)}  \right)\right\vert_{\D \rightarrow 0}\right\vert_{m\rightarrow0},
\end{eqnarray}
where in the second line, one cannot switch the order of derivatives trivially.

\subsection{Krylov entropy and baby universe}
\begin{figure}[h!]
    \centering
\begin{tikzpicture}
\node[] () at (0,0)
    {\centering \includegraphics[width=0.3\textwidth]{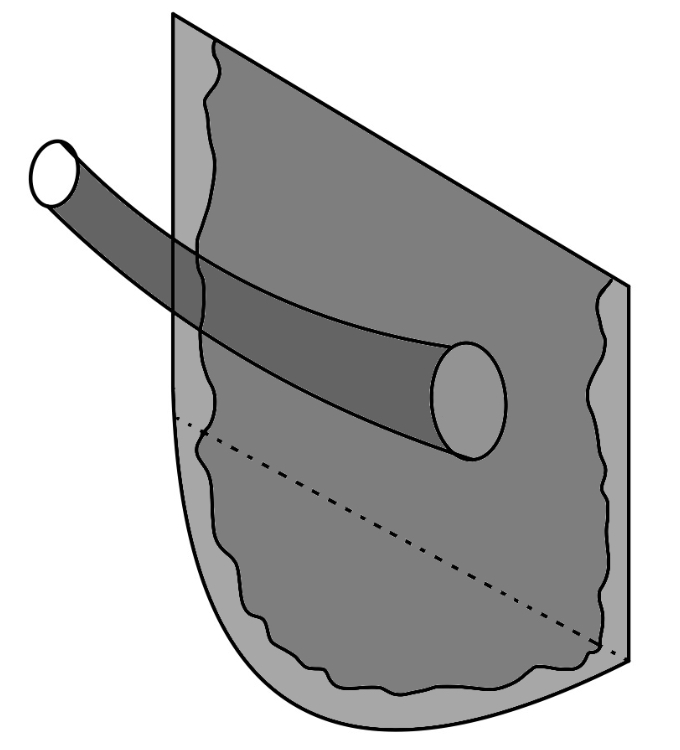}};
    \node[]  at (0.5,2) {$L$};
\node[]  at (-2.3,1.5) {$\mathbf{b}$};
\node[]  at (2.9,-2.5) {$\mathbf{\ket{L=0}}_{\beta}\otimes \ket{0}$};
\node[]  at (2.5,1.3) {$\mathbf{\bra{L}\otimes \bra{b}}$};
\node[]  at (2.9,-0.5) {$t/2$};
\draw[->][line width=0.4mm](2.5,-2)to(2.5,0.5);
\end{tikzpicture}
\caption{An illustration of the wavefunction $\psi_{L,b}(t)$ in Eq. \eqref{EQ377}: the parent universe emits a baby universe of size \texttt{b} while its geodesic boundary length $L$ evolves. The Euclidean half-disk at the bottom sets the Hartle-Hawking state, time evolution propagates it by $t/2$, the geodesic loop of length \texttt{b} marks the baby universe, and the geodesic boundary of length $L$ is the remaining parent.}
\label{fig-Wavefunction}
\end{figure}

For Krylov complexity in quantum systems, one can define an associated quantity called Krylov entropy \cite{Barbon:2019wsy}:
\begin{equation}
    S(t)=-\sum_{n}|\varphi_n(t)|^2 \log(|\varphi_n(t)|^2) ,
\end{equation}
where $\varphi_n(t)$ is the transition amplitude:
\begin{equation}
    \varphi_n(t)=\braket{n}{\varphi(t)}=\bra{n}e^{-i\hat{T}t} \ket{\varphi(0)}.
\end{equation}
In the limit as $t \rightarrow 0$, the state is localized, resulting in a Krylov entropy of zero. In the limit $t \rightarrow \infty$, if the system is chaotic enough, and the chord basis is of dimension $N$, we have $ |\varphi_n(t)|^2=\frac{1}{N}$. Thus, the Krylov entropy is $S(\infty)=-\sum_L 1/N \log(1/N)=\log N$, which indicates that the state becomes delocalized in the geodesic basis after time evolution. Thus, the range of Krylov entropy is 
\begin{equation}
    0 \leq S(t) \leq \log N.
\end{equation}
In disorder-averaged theories like SYK, one should look at the ensemble-averaged density matrix: 
\begin{equation}
    \overline{\rho(t)}=\int dH P(H) e^{-i\hat{H}t}\ket{\varphi(0)}\bra{\varphi(0)} e^{i\hat{H}t},
\end{equation}
which stays pure for any $t$ with or without ensemble averages. In the DSSYK, the ensemble averaging is encoded in the chord Hilbert space formalism, and the density matrix for the time-evolved initial state is
\begin{equation}
    \rho(t)= e^{-i\hat{T}t}\ket{\varphi(0)}\bra{\varphi(0)} e^{i\hat{T}t}.
\end{equation}
This shall give a trivial von Neumann entropy because the state from the boundary observer's point of view is always pure. This is the consequence of the lack of detectability of information in the bulk from the boundary \cite{Penington:2023dql}. However, in the holographic dual picture of gravity, this would change if one studies the wormhole in the open universe with baby universes. In this setup, we try to write the Krylov entropy in the bulk Hilbert space. This can be done through the so-called third-quantization scheme \cite{Giddings:1988wv, Marolf:2020xie, Penington:2023dql},\footnote{See also \cite{Aguilar-Gutierrez:2025hty} for a slightly different approach.} which has already been explored in JT gravity. Although baby universes are not rigorously studied yet in sine-dilaton gravity, it is worthwhile pointing out that our discussions here do not depend on details of the theory. We consider the total bulk Hilbert space as 
\begin{equation}
\mathcal{H}_{\mathrm{tot}}=\mathcal{H}_{\mathrm{geo}}\otimes \mathcal{H}_{\mathrm{baby}},
\end{equation}
with state being written as $\ket{L}\otimes \ket{b}$. This factorization should be understood purely in the context of quantum gravity once wormhole/baby-universe effects are included. Our goal is to identify the correct bulk dual of the boundary Krylov basis. Before tracing/factorizing out the baby-universe sector, the naive geodesic basis is not the appropriate orthonormal basis (in particular, wormhole
effects lead to the ``wormhole shortening'' discussed above). After tracing/factorizing out
the baby-universe sector, the geodesic basis provides an orthonormal basis that can be
identified with the bulk representation of the Krylov basis. In this framework, the boundary
Krylov entropy admits a natural bulk interpretation as the entanglement entropy between
the geometric sector and the baby-universe sector.\footnote{This should not be interpreted as a factorization of the boundary Krylov basis: boundary
observers are blind to the baby-universe sector.}
Therefore, the initial state can be written as:
\begin{equation}
    \ket{\varphi(0)}=\ket{L=0}\otimes\ket{0}_{\mathrm{baby}}.
\end{equation}
After time evolution, the state becomes
\begin{equation}
    \ket{\psi(t)}=\sum_{L,b}\psi_{L,b}(t) \ket{L}\otimes \ket{b},
\end{equation}
where this wavefunction $\psi_{L,b}(t)$ is the propagator $P_{tr}$ from \cite{Saad:2019pqd} setting one boundary as $L=0$ for nonzero $b$. It can be written as follows in JT gravity\footnote{The geodesic in JT gravity is defined as $L_{\text{JT}}=\int ds$, while in sine-dilaton gravity, it is defined as $L_{\text{SD}}=\int ds e^{-i\Phi/2 }$ \cite{Blommaert:2024ymv}.}:
\begin{eqnarray}\label{EQ377}
    \psi_{L,b}(t)&=&(\bra{L}\otimes\bra{b})e^{-it H_{JT}}(\ket{L=0}_{\beta}\otimes \ket{0})\nonumber\\ &=&\int^\infty_0 dE \frac{16\cos(b\sqrt{2E})}{\pi \sqrt{2E}}e^{-i(t-i\beta/2)E}e^{-L/2}K_{i\sqrt{8E}}(4e^{-L/2})K_{i\sqrt{8E}}(4),
\end{eqnarray}
which can be obtained from the trumpet wavefunctions. The summation of $b$ is from $0$ to $\infty$. When $b\rightarrow 0$, $\psi_{L,0}(t)\rightarrow \psi_{L}^{\text{Disk}}(t)$. It is shown in the figure \ref{fig-Wavefunction} in the piecewise Euclidean and Lorentzian geometry. The initial state $\ket{L=0}$ has a constant overlap with bulk energy eigenstates: $\braket{E}{L=0}=1$, which indicates that the initial state is the maximally entangled state.\footnote{This initial state can also be understood in the Euclidean disk with defect. $L=0$ is the minimal geodesic that wraps around once the half-defect \cite{Blommaert:2024ymv}.} After the Euclidean preparation, it becomes the TFD state, which should be considered as dual to a Hartle-Hawking state $\ket{HH_\beta}$ in the presence of a proper ensemble-averaged quantum theory \cite{Saad:2019pqd}.\footnote{It has been shown that the path integral over wormhole topologies yields non-trivial late-time contributions, establishing a connection with random matrix behavior and the ETH interpretation~\cite{Saad:2019pqd}.} Therefore, the wavefunction $\psi_{L,b}(t)$ is proportional to the trumpet wavefunction in this setup. Note that we omit contributions of emitting and reabsorbing baby universes, which are suppressed by the factor $e^{-2nS_0}$ with $\chi=1-2n$ from topology changes. While in sine-dilaton gravity, given the trumpet partition function from \cite{Blommaert:2025avl}, we are able to derive the trumpet wavefunction $\psi^{\text{SD}}_{\text{Tr},\beta/2}(L,b)$ as
\begin{equation}
    \psi^{\text{SD}}_{\text{Tr},\beta/2}(L,b)=\int d\theta \frac{\cos(b\theta)}{(q^2,q^2,e^{\pm i\theta};q^2)_{\infty} }  e^{-\frac{\beta}{2}E(\theta)} \psi_{\theta}(L),
    \label{eq-trumpetwaveSD}
\end{equation}
where $E(\theta)$ is the spectrum of sine-dilaton gravity given below (\ref{eq-wavefunctionSD}) and the $\psi_{\theta}(L)$ is the wavefunction in (\ref{eq-wavefunctionSD}). Detailed derivations of trumpet wavefunctions in JT and sine-dilaton gravity are given in the Appendix \ref{sec-appA}.


The density matrix of the time-evolved initial state can be written in this picture as:
\begin{equation}
    \rho_{\mathrm{tot}}(t)=\ket{\psi(t)} \bra{\psi(t)}=\sum_{L,b;L',b'}\psi_{L,b}(t)\psi^*_{L',b'}(t) \ket{L}\bra{L'}\otimes \ket{b}\bra{b'} \,.
\end{equation}
Note that at this point, the density matrix remains pure since the closed system unitary evolution does not change the purity.
Consider that they both form an orthogonal basis\footnote{The orthogonality condition is only required at the disk level in the third quantization picture. Although it is naturally satisfied in the canonical ensemble, it fails to be orthogonal even at the disk level in the microcanonical ensemble \cite{Miyaji:2024ity}. Nevertheless, one can force it through the Gram-Schmidt process, which is actually the natural steps in the Lanczos algorithm. One can thus always interpret $p_L$ as a probability.}:
\begin{equation}
    \braket{L}{L'}\propto\delta(L-L'),\hspace{0.8cm} \braket{b}{b'}\propto\delta_{bb'}.
\end{equation}
One can trace out the baby universe part and be left with the density matrix only from the geodesic basis:
\begin{equation}
    \rho_{\mathrm{geo}}(t)=\Tr_{\mathrm{baby}}(\rho_{\mathrm{tot}}(t))=\sum_{L}\vert \psi_{L}(t)\vert^2 \ket{L}\bra{L},
\end{equation}
where $\rho_{\mathrm{geo}}(t)$ now becomes a diagonal mixed density matrix since the geodesic basis is diagonal. Here, $\psi_{L}(t)$ shall be dual to the transition amplitudes shown in Krylov entropy in DSSYK. This is because we can only have an orthonormal geodesic length basis after separating the baby universe sector, which can serve as a bulk Krylov basis.
We can thus find the von Neumann entropy for this density matrix as 
\begin{equation}
    -\Tr(\rho_{\mathrm{geo}}(t) \log \rho_{\mathrm{geo}}(t))=-\sum_L\vert \psi_{L}(t)\vert^2 \log \vert \psi_{L}(t)\vert^2 \longleftrightarrow  S(t).
\end{equation}
We argue that in the Hilbert space including baby universe, \textit{the original Krylov entropy, defined as the Shannon entropy, becomes the von Neumann entropy of $\rho_{\mathrm{geo}}(t)$ in the bulk}. This von Neumann entropy quantifies the amount of information lost to the baby universe as the initial state gets increasingly entangled with it over time. This picture involving the baby universe is consistent with the ensemble average interpretation: the presence of different ensembles can be understood as arising from the existence of baby universes. Since, overall, the baby universe is formally defined in the gravitational path integral, which is an equivalent interpretation of the matrix ensemble integral \cite{Marolf:2020xie}.

%
\section{Conclusions}\label{sec-conclude}
\begin{table}[t!]
    \centering
    \renewcommand{\arraystretch}{1.2}
    \setlength{\tabcolsep}{10pt}
    \begin{tabular}{ll}
    \toprule
        \textbf{(Boundary) Krylov Observable} & \textbf{(Bulk) Holographic Gravity Dual} \\ 
        \midrule
        \midrule
        Krylov complexity & Wormhole length \\
        \midrule
        Growth rate of Krylov complexity & Wormhole velocity; \\
        & Proper momentum (small-$E$ or early-$t$ regime) \\
        \midrule
        Higher-order Krylov complexities & R\'{e}nyi-like wormhole length \\
        \midrule
        Logarithmic Krylov complexity & Probe wormhole length via replica wormholes \\
        \midrule
        Krylov entropy & Entanglement entropy of geodesic \\
        & and baby-universe sectors \\
\bottomrule
    \end{tabular}
    \caption{Krylov-space observables in the double-scaled SYK model and their dual descriptions in sine-dilaton/JT gravity.}
    \label{tab-summary}
\end{table}

In this work, we extended the program to establish holographic duals for Krylov-space observables in the double-scaled Sachdev-Ye-Kitaev (DSSYK) model. Building on the known duality between Krylov complexity and geodesic length in Jackiw-Teitelboim (JT) and sine-dilaton gravity, we identified the growth rate of Krylov complexity as dual to the wormhole velocity, and clarified a complementary interpretation in terms of the proper momentum of an infalling particle in the semiclassical regime and at early times. Furthermore, by studying expectation values in coherent states, we showed that this observable can serve as a boundary diagnostic of bulk features such as firewalls.

More specifically, motivated by the proposal that negative eigenvalues of the wormhole velocity operator signal a white-hole/firewall, we analyzed the dual boundary quantity, the growth rate of Krylov complexity in DSSYK, focusing on the coherent-state basis. We showed that these eigenvalues are determined by the imaginary parts of the coherent-state parameters. Using the HKLL bulk reconstruction method, we found that indications of a firewall are encoded in the negative eigenvalues.

We then extended the dictionary to more refined quantities in the Krylov subspace. We argued that higher-order Krylov complexities naturally map onto connected replica wormholes, providing a bulk counterpart to R\'enyi-like refinements of operator growth. The logarithmic Krylov complexity is defined via the replica trick and probes wormhole saddles beyond the disconnected phase. Finally, we studied Krylov entropy, defined as the Shannon entropy of transition amplitudes. On the bulk side, we found that it equals the von Neumann entropy of the reduced density matrix obtained by tracing out baby universes in the third-quantization picture.\footnote{See \cite{Aguilar-Gutierrez:2025otq} for a different discussion on the von Neumann entropy of subsystems in DSSYK and Krylov complexity} Thus, the bulk dual of Krylov entropy quantifies the entanglement between the geodesic-length basis and the baby-universe basis, thereby measuring the information lost to the baby-universe sector.

The holographic dualities discussed in this paper are summarized in Table~\ref{tab-summary}. This summary highlights the utility of the Krylov subspace as a boundary framework for diagnosing not only quantum chaos but also holographic duality. More broadly, the developments here reinforce the view that Krylov-based methods are powerful probes of operator growth and state spreading, and provide windows into the emergent structure of spacetime.

Several directions remain for future exploration. It would be valuable to test these dualities in other solvable holographic models, to study the role of coherent states and quantum information geometry more systematically, and to investigate possible signatures of baby universes in boundary observables. It is also natural to ask whether this Krylov-based holographic picture continues to hold under quantum corrections, order by order; see \cite{Bossi:2024ffa,Blommaert:2025avl,Okuyama:2025fhi} for recent developments.\footnote{In Section~\ref{sec-Coherent DSSYK}, we derive the coherent states in DSSYK and compute the growth rate of Krylov complexity in the coherent-state basis. It is straightforward to compute the Fubini-Study metric for this q-deformed oscillator. It would be interesting to test whether the statement that Krylov complexity equals volume in the information metric continues to hold in the $q$-deformed algebra.}

Exploring the proposed dualities in higher dimensions is a long-term, technically challenging objective. The simplest extension may lie in the AdS$_3$/CFT$_2$ correspondence. Even in two dimensions, given the established correspondence between observables in DSSYK and sine-dilaton gravity, a natural question arises: \textit{Is this duality unique?} More concretely, starting from the Krylov complexity of DSSYK, one may ask whether analogous quantities in other two-dimensional dilaton gravity models with different dilaton potentials and boundary conditions reproduce the same behavior. A notable example is the dS limit of sine-dilaton gravity, recently explored in \cite{Heller:2025ddj}.\footnote{Another related issue has been addressed in JT gravity \cite{Stanford:2019vob}.} 

Another accessible direction is to examine different initial states, for instance, the thermofield-double (TFD) state expressed in the chord basis, which requires only the construction of the corresponding Krylov basis. We plan to report further progress on some of these directions in the near future.

%
\acknowledgments
We would like to thank {Sergio E. Aguilar-Gutierrez, Kuntal Pal, Jaydeep Kumar Basak, Cheng Peng and Zhenbin Yang} for valuable discussions and correspondence.
This work was supported by the Basic Science Research Program through the National Research Foundation of Korea (NRF) funded by the Ministry of Science, ICT \& Future Planning (NRF-2021R1A2C1006791), the Korea government(MSIT)(RS-2025-02311201), (RS-2024-00445164) and the framework of international cooperation program managed by the NRF of Korea (RS-2025-02307394), the Creation of the Quantum Information Science R\& D Ecosystem (Grant No. RS-2023-NR068116) through the National Research Foundation of Korea (NRF) funded by the Korean government (Ministry of Science and ICT). 
This research was also supported by GIST research fund (Future leading Specialized Resarch Project, 2026, and the Regional Innovation System \& Education(RISE) program through the (Gwangju RISE Center), funded by the Ministry of Education(MOE) and the (Gwangju Metropolitan City), Republic of Korea.(2025-RISE-05-001).
HSJ was supported by an appointment to the JRG Program at the APCTP through the Science and Technology Promotion Fund and Lottery Fund of the Korean Government. HSJ was also supported by the Korean Local Governments -- Gyeongsangbuk-do Province and Pohang City.
HSJ and JFP are supported by the Spanish MINECO ‘Centro de Excelencia Severo Ochoa' program under grant SEV-2012-0249, the Comunidad de Madrid ‘Atracci\'on de Talento’ program (ATCAM) grant 2020-T1/TIC-20495, the Spanish Research Agency via grants CEX2020-001007-S and PID2021-123017NB-I00, funded by MCIN/AEI/10.13039/501100011033, and ERDF A way of making Europe.
YF also thanks Cheng Peng and the Kavli Institute for Theoretical Sciences (KITS) at the University of Chinese Academy of Sciences (UCAS) for hospitality during the final stages of this project.
All authors contributed equally to this paper and should be considered as co-first authors.

%
\appendix
\section{Trumpet wavefunctions $\psi_{\mathrm{Tr},\beta/2}(L,b)$ in JT and sine-dilaton gravity}\label{sec-appA}
In JT gravity, the wavefunction of canonical Hamiltonian in the geodesic basis is
\begin{equation}
    \psi_E(L)=4e^{-L/2}K_{i\sqrt{8E}}(4e^{-L/2}),
\end{equation}
where $K_n$ is the modified Bessel function of the second kind. This wavefunction obeys the orthogonality condition:
\begin{equation}
    e^{-S_0}\int dL e^L \psi^*_E(L)\psi_{E'}(L)=\frac{\delta(E-E')}{\rho_0(E)},
    \label{eqAPP-orthogonality}
\end{equation}
where $\rho_0(E)=e^{S_0} \frac{\sinh(2\pi \sqrt{2E})}{2\pi^2}$ is the density of states in JT gravity. Now, consider a single trumpet partition function from the Euclidean path integral
\begin{eqnarray}
    Z_{\mathrm{Tr}}(\beta,b)&=&e^{-S_0}\int e^L \braket{HH_\beta}{L} \braket{L,b}{HH_\beta}dL\nonumber\\
    &=& e^{-S_0}\int e^L \psi^*_{D,\beta/2}(L) \psi_{\mathrm{Tr},\beta/2}(L,b) dL,
    \label{eqApp-partitionTr}
\end{eqnarray}
where $b$ is the size of the baby universe and $\psi^*_{D,\beta/2}(L)$ is the disk wavefunction given as 
\begin{equation}
    \psi^*_{D,\beta/2}(L)\equiv \braket{HH_\beta}{L}=\int^\infty_0 dE \rho_0(E) e^{-\beta E/2}\psi^*_E(L).
    \label{eqApp-wavefuncD}
\end{equation}
Consider the trumpet wavefunction can be written in the same form:
\begin{equation}
    \psi_{\mathrm{Tr},\beta/2}(L,b)=\int^\infty_0 dE \rho_{\mathrm{Tr}}(E,b) e^{-\beta E/2}\psi_E(L).
    \label{eqApp-wavefuncTr}
\end{equation}
Therefore, the task is to find the trumpet density $\rho_{\mathrm{Tr}}(E,b)$. In order to do so, we plug (\ref{eqApp-wavefuncTr}) and (\ref{eqApp-wavefuncD}) into (\ref{eqApp-partitionTr}) and integrate over $L$. By using the orthogonality condition (\ref{eqAPP-orthogonality}), one can get
\begin{equation}
    Z_{\mathrm{Tr}}(\beta,b)=\int^\infty_0 dE \rho_{\mathrm{Tr}}(E,b) e^{-\beta E}.
\end{equation}
The trumpet partition function is given in \cite{Saad:2019lba} as 
\begin{equation}
    Z_{\mathrm{Tr}}(\beta,b)=\frac{1}{\sqrt{2\pi \beta}}e^{-\frac{b^2}{2\beta}},
\end{equation}
where we take $\gamma=1$. To derive the trumpet density, one simply employs the inverse Laplace transform
\begin{equation}
    \mathfrak{L}^{-1}\left[ \frac{1}{\sqrt{x}}e^{\a x}\right](y)=\frac{\cosh(2\sqrt{\a y})}{\sqrt{\pi y}}.
\end{equation}
One can straightforwardly obtain the trumpet density as
\begin{equation}
    \rho_{\mathrm{Tr}}(E,b)=\frac{\cos(b\sqrt{2E})}{\pi \sqrt{2E}}.
\end{equation}
Thus, the trumpet wavefunction can be written as \cite{Saad:2019pqd} 
\begin{equation}
    \psi_{\mathrm{Tr},\beta/2}(L,b)= \int^\infty_0 dE \frac{\cos(b\sqrt{2E})}{\pi \sqrt{2E}} e^{-\beta E/2}\psi_E(L).
\end{equation}
In sine-dilaton gravity, unlike JT gravity, the geodesic is intrinsically discretized. One therefore should not consider $L$ as the same quantity in two theories. Instead of the energy basis, it is more convenient to use the $\theta$-basis, where the wavefunction with respect to geodesic length is given in (\ref{eq-wavefunctionSD}), which is written in terms of q-Hermite polynomials. They are known to obey the $q$-Mehler formula
\begin{equation}
    \sum^\infty_n \frac{H_n(\cos\theta|q^2)H_n(\cos\theta'|q^2)}{(q^2;q^2)_\infty}r^n=\frac{(r^2;q^2)_\infty}{(re^{\pm i(\theta\pm\theta')};q^2 )_\infty}.
\end{equation}
In the limit of $r\rightarrow 1$, the right-hand side can be approximated by $(q^2;q^2)_\infty \delta(\theta-\theta')$. Therefore,
\begin{equation}
    \sum^\infty_n \frac{H_n(\cos\theta|q^2)H_n(\cos\theta'|q^2)}{(q^2;q^2)_\infty}\approx (q^2;q^2)_\infty \delta(\theta-\theta').
\end{equation}
Similarly to JT gravity, by using the above formula, we can write the trumpet partition function in sine-dilaton gravity as
\begin{equation}
    Z^{\mathrm{SD}}_{\mathrm{Tr},\beta/2}=\int^\pi_0 d\theta \rho_{\mathrm{Tr}}(\theta,b)(q^2,q^2,e^{\pm2i\theta};q^2)_\infty e^{-\beta E(\theta)}.
    \label{eqApp-partitionSD}
\end{equation}
On the RHS, the integrand is a periodic function in $\theta$, and we introduce
\begin{equation}
    \tilde{\beta}\equiv \frac{\beta}{2|\log q|},
\end{equation}
and define
\begin{equation}
    f(\theta,b)\equiv \rho_{\mathrm{Tr}}(\theta,b)(q^2,q^2,e^{\pm2i\theta};q^2)_\infty.
\end{equation}
We then expand the function $f$ in terms of its Fourier modes:
\begin{equation}
   f(\theta,b)= \sum_{n=0}^\infty c_n(b) \cos(n \theta),
\end{equation}
where the odd part of the Fourier modes vanishes in (\ref{eqApp-partitionSD}). We can therefore rewrite the trumpet partition function as
\begin{equation}
    Z^{\mathrm{SD}}_{\mathrm{Tr}}(\beta,b)=\int^\pi_0 d\theta e^{-\tilde{\beta}\cos(\theta)} \sum_{n=0}^\infty c_n(b) \cos(n \theta).
\end{equation}
Consider the integral representation of the Modified Bessel function of the first kind
\begin{equation}
    I_n(x)=\frac{1}{\pi}\int^\pi_0 e^{x \cos y}\cos(n y)dy,
\end{equation}
we can find
\begin{equation}
Z^{\mathrm{SD}}_{\mathrm{Tr},\beta/2}=\sum_{n=0}^\infty c_n(b) I_n(\tilde{\beta}).
\label{eqAPP-parSDmodes}
\end{equation}
\cite{Blommaert:2025avl} derived the trumpet partition function in sine-dilaton gravity as
\begin{equation}
    Z^{\mathrm{SD}}_{\mathrm{Tr}}(\beta,b)=I_b(\tilde{\beta}).
\end{equation}
Comparing with (\ref{eqAPP-parSDmodes}), we find the coefficients as
\begin{equation}
    c_n(b)=\delta_{nb},
\end{equation}
which leads to the following expression for trumpet density:
\begin{equation}
    \rho_{\mathrm{Tr}}(\theta,b)=\frac{\cos(b\theta)}{(q^2,q^2,e^{\pm2i\theta};q^2)_\infty}.
\end{equation}
Therefore, the trumpet wavefunction in sine-dilaton gravity is given as
\begin{equation}
    \psi^{\mathrm{SD}}_{\mathrm{Tr},\beta/2}(L,b)=\int d\theta \frac{\cos(b\theta)}{(q^2,q^2,e^{\pm2i\theta};q^2)_\infty} e^{-\beta E(\theta)/2} \psi_\theta(L),
\end{equation}
which is exactly (\ref{eq-trumpetwaveSD}).

\bibliography{Refs}
\bibliographystyle{JHEP}

\end{document}